\documentclass[aps,twocolumn]{revtex4}
\usepackage{graphicx}

\input epsf

\def\be{\begin{equation}}
\def\ee{\end{equation}}
\def\bea{\begin{eqnarray}}
\def\eea{\end{eqnarray}}

\def\TeV{\,{\rm TeV}}

\def\pc{\,{\rm pc}}

\def\mpc{\,{\rm Mpc}}

\def\cmm2{{\,\rm cm^{-2}}}
\def\cm2{{\,{\rm cm}^2}}
\def\cmm3{{\,{\rm cm}^{-3}}}
\def\gcmm3{{\,{\rm g\,cm^{-3}}}}
\def\kms{\,{\rm km\,s^{-1}}}

\def\fun#1#2{\lower3.6pt\vbox{\baselineskip0pt\lineskip.9pt
  \ialign{$\mathsurround=0pt#1\hfil##\hfil$\crcr#2\crcr\sim\crcr}}}

\def\vec{\bf}

\def\kms{{\rm~km~s^{-1}}}

\def\mpc{{\rm~Mpc}}
\def\msun{{\,M_\odot}}

\def\etal{{\it et al. }}

\def\p3m{P$^3$M}

\def\la{\mathrel{\mathpalette\fun <}}
\def\ga{\mathrel{\mathpalette\fun >}}
\def\fun#1#2{\lower3.6pt\vbox{\baselineskip0pt\lineskip.9pt
  \ialign{$\mathsurround=0pt#1\hfil##\hfil$\crcr#2\crcr\sim\crcr}}}

\begin{document}
\title{Dark Matter from Early Decays}
\author{Manoj\ Kaplinghat}
\affiliation{Department of Physics and Astronomy, 4129 Frederick
  Reines Hall\\
University of California, Irvine, California 92697, USA}
\date{\today}

\begin{abstract}
Two leading dark matter candidates from supersymmetry and other
theories of physics beyond the standard model are WIMPs and
weak scale gravitinos. If the lightest stable particle is a gravitino,
then a WIMP will decay into it with a natural lifetime of order a
month $\sim M_{\rm pl}^2/M_{\rm weak}^3$. We show that if the bulk of
dark matter today came from decays of neutral particles with 
lifetimes of order a year or smaller, then it could lead to a
reduction in the amount of small scale substructure, less concentrated 
halos and constant density cores in the smallest mass halos. Such
beneficial effects may therefore be realized naturally, as discussed
by Cembranos, Feng, Rajaraman, and Takayama, in the case of
supersymmetry.
\end{abstract}
 \maketitle

\section{Introduction}

We have many independent lines of observational evidence that point to
the existence of dark matter in the universe. The dark matter
particle, however, has not been observed.  Most theories of
physics beyond the standard model that have new physics at the weak
scale, naturally predict the existence of WIMP (Weakly Interacting
Massive Particle) dark matter. The weak interaction cross-section
endows the WIMP with the right relic density to make up a substantial
fraction of the energy density of the universe. Thus most cosmological
studies of dark matter assume that it is a stable cold particle with
interaction  cross-sections comparable to the weak interaction
cross-sections. 

Within theories of physics beyond the standard model, one also
has the gravitino (or other gravitationally coupled particles). Recent work
by Feng, Rajaraman and Takayama \cite{feng03a} provided an 
interesting twist to the old dark matter tale. They observed that if
the gravitino is the lightest supersymmteric particle (LSP), then the 
next lightest particle (NLSP) can decay into the gravitino with a
lifetime that is long since the gravitino couples only
gravitationally. The natural time scale for this decay is $M_{\rm
  pl}^2/M_{\rm weak}^3 \sim\,$ month. This opens up the
possibility that at least a fraction of the dark matter we observe
today comes from decays when the universe was days to years old. We
note that similar models may be constructed in the context of
cosmologies with extra dimensions \cite{feng03c}. Two early universe
constraints on these models are that the electromagnetic decay
products must not distort the black body spectrum appreciably and the
light element abundances from BBN must match with observations. Recent
work has shown that there are regions of parameter space where all
these conditions may be met
\cite{feng03b,feng04a,feng04b,roszkowski04,lamon05}. In the 
present work we work out the late time consequences of these decays. 

Another way to get dark matter to decay around similar epochs is to
kinematically suppress the decays such that the lifetime is orders of
magnitude larger than the ``natural'' time scale. Profumo
\etal \cite{profumo04} provided the first example
of such a decay that arises in supersymmetric theories. The consider
the decay of stau into the LSP with a mass splitting smaller than the
tau lepton mass. In this case, the two body decay is forbidden and the
stau can only decay into 4 daughter particles. The 4-body decay of a
charged slepton is  suppressed by the NLSP-LSP mass splitting as
$(M/\Delta M)^8$. Thus with $\Delta M/M \sim 10^{-3}$, the stau
lifetime is of order a few weeks \cite{profumo04}. 

Sigurdson and Kamionkowski \cite{sigurdson03} and then Profumo \etal
\cite{profumo04} worked out the impact of the 
charged stau decay on the matter and CMB power spectrum. They found
that the dark matter power spectrum was suppressed due to the coupling
of the charged NLSP to the photon-baryon system. This coupling damps the
matter power spectrum once a perturbation mode enters the horizon. 

In this paper we point out that the large suppression of the matter
power spectrum is possible {\em even if} the NLSP is not
charged. Unless the mass splitting between the NLSP and LSP is
fine-tuned to be small, we point out that the decay will impart a
large velocity to the LSP. Effectively the LSP will act like warm dark
matter. This leads to a cut-off in the matter power spectrum. We
calculate shape of the power spectrum given the lifetime and the
masses of the particles involved.  

We also calculate the effect of the smaller phase space density that
results from this decay on dark matter halos. They could (for small
enough primordial phase space density) be instrumental in shaping the
smallest mass halos, lowering the fraction of mass in sub-halos, and
reducing the concentration of larger mass halos. Stringent constraints
on these decays come from the requirement that the universe be
reionized by a redshift of six or higher. If the present hints
\cite{kogut03} for early reionization are borne out by future data,
then this could rule out a big chunk of the supersymmetric parameter
space in which the gravitino is the lightest particle. 

Cembranos \etal \cite{cembranos05} show that the beneficial effects
mentioned above may be obtained in a large region of the
super-symmetric parameter space.  The prospects for testing this
region of parameter space wherein we have early decays is
promising. These models will be tested by observations made with NGST
of the high redshift universe (e.g., \cite{mesinger05}), by future
surveys of strong lens systems (e.g., \cite{dalal02}), weak lensing
measurements (e.g., \cite{dolney04}), as well as  the more traditional
CMB and large scale structure observations.

The dark matter observed today could also be a mixture of particles
produced during the reheating phase and those arising from the
decays. For the gravitino example, the two contributions are expected
to be comparable for a reheating temperature of $10^{10}$ GeV. This
kind of mixed model would ameliorate the rapid drop in power on small
scales. 

Dark matter from decays have a non-thermal momentum distribution. Lin
\etal \cite{lin00}, Hisano \etal \cite{hisano00}, Kitano and Low
\cite{kitano05} studied different models of non-thermal WIMP dark
matter and their effect on the small scale structure. The generic
cosmological effects they found are similar to what we discuss here
because the effects stem from the large velocity dispersion. Decays
leading to a large velocity dispersion were also considered briefly by
Hogan and Dalcanton \cite{hogan00} in their work on the astrophysical 
consequences of a small primordial phase space density. 

\section{Dark matter from decays}

\newcommand{\ddm}{\mathrm{DDM}}
\newcommand{\dm}{\mathrm{DM}}
\newcommand{\rd}{\mathrm{L}}
\newcommand{\pcm}{p_{\mathrm CM}}
\newcommand{\adec}{a_{\rm dec}}
\newcommand{\adot}{{\dot a}}
\newcommand{\dfdm}{\delta \! f_\mathrm{DM,\ell}}
\newcommand{\dfddm}{\delta \! f_\mathrm{DDM,\ell}}
\newcommand{\erf}{\mathrm{Erf}}
\newcommand{\fsl}{\lambda_{fs}}

Consider the decay process, DDM $\rightarrow$ DM + L,
where L denotes ``light'' particles with mass $m_1$, DM is the dark
matter today with mass $m$ and DDM is the parent particle with mass
$M$. When we derive the perturbations, we will be enforce the limit
$m_1 << m$ as is appropriate for supersymmetric models (e.g.,
\cite{feng04a}).  
We will also assume that DDM and DM are neutral, for example, a
sneutrino decaying into a neutrino and a gravitino. All of the effects
we mention here are relevant also for charged particle decay. In
addition, the coupling of the charged parent particle to the
photon-baryon system results in the damping of perturbations that come
into the horizon when the charged parent particles dominate the
non-relativistic matter density \cite{sigurdson03}. We will work out 
the physics and astrophysics of charged particle decay in a future
publication. 

For a generic n-body decay, neglecting Pauli-blocking factors and
inverse decays, the change in the phase space due to decays is
given by
\begin{equation}
\dot{f}_\ddm(p_\ddm) = - {a M \over E_\ddm \tau} f_\ddm(p_\ddm)\,,
\label{eq:fddmdot} 
\end{equation}
where $E_\ddm^2=p_\ddm^2+M^2$ and $f_\ddm$ is the phase space
distribution of the DDM particles. The over-dot denotes derivative
with respect to the conformal time, $d\eta = dt/a$ and $a$ is the scale
factor. Specializing to two-body decays, one can show that the DM
phase space is populated by the DDM decays according to the equation
\cite{kawasaki92, kang93}: 
\begin{equation}
\dot{f}_\dm(p_\dm) = {a M^2 \over 2 \tau E_\dm p_\dm \pcm} \int_{E_1}^{E_2}
dE f_\ddm(p) \,, \label{eq:fdmdot}
\end{equation}
where $E_{2,1} = (0.5E_\dm m_0^2 \pm p_\dm \pcm M)/m_\dm^2 $. $\pcm$
is the center-of-mass momentum and $m_0^2 \equiv
M^2+m_\dm^2-m_\rd^2$. An analogous equation holds for the other
daughter particle. Note that for a two-body decay we have lifetime 
$\tau=8\pi M^2/\pcm|{\cal M}|^2$ where $|{\cal M}|^2$ is the quantum
mechanical amplitude for the decay process. Also note that since the
decay happens when the DDM particle is cold, $\pcm$ is an accurate
measure of the momentum imparted to the daughter particles. 
 
The equations for the change in the phase space distribution
due to the decays, Eq.~\ref{eq:fddmdot} and Eq.~\ref{eq:fdmdot},  can be
integrated to yield  the following equations for the change in the
density of DM and DDM.  
\begin{eqnarray}
\dot{\rho}_\ddm + 3 {\adot \over a}\rho_\ddm &=& - {a \over \tau} \rho_\ddm \,,
\label{eq:rhoddmdot}\\ 
\dot{\rho}_\dm + 3 {\adot \over a} ( \rho_\dm + P_\dm ) &=& {a m_0^2
  \over 2 \tau M^2} \rho_\ddm \,, \label{eq:rhodmdot}
\end{eqnarray}
where $P_\dm$ is the pressure of DM and it is comparable to
$\rho_\dm$ at early times when the bulk of decay occurs. An equation
for ``L'' particles may be written down by inspection of the DM
equation. 

We will also need to calculate the phase space distribution of DM
particles. One can show that this is (in the limit of
completely non-relativistic decay) given by
\begin{equation}
f_\dm(q,a)  =  { (2\pi)^2 \Omega_M \rho_\mathrm{crit} t_q \over m q^3
  \tau } \exp(-{t_q \over \tau}) \Theta(a\pcm -q) \,,\label{eq:fdm}
\end{equation}
where $q$ is the comoving momentum of the DM particle and
$t_q=t(a=q/\pcm)$. We have assumed that the decays are happening
during radiation domination and hence $t_q \propto q^2$ and $f_\dm(q)
\propto q^{-1}\exp(-q^2/[2\pcm^2\tau \adot])$.  


\section{Perturbations}

We now write down the equations for the perturbations in DDM, DM
and L particles. The first result is that the
perturbation in density relative to the mean for the
DDM particles is unchanged. This is derived by using the fact that the
DDM particles are  non-relativistic. The physical content of this
statement is that since the DDM particles do not have a large peculiar
velocity during decay, the only effect of the decay is to remove the
same fraction of the DDM particles from every region of space.  

To calculate the perturbations explicitly, we write it as a sum of two
terms: 
\begin{equation}
\dfdm(k,q,a) = \dfdm^{(a)}(k,q,a) + \dfdm^{(b)}(k,q,a)
\,.\label{eq:dfdm}  
\end{equation}
The first term $\dfdm^{(a)}(k,q,a)$ is chosen to satisfy the
collisionless Boltzmann equation for a massive particle and may be
calculated in a manner analogous to the perturbations in the massive
neutrino phase  space \cite{ma95}. The second term is harder to
calculate numerically. One can make approximations similar to the one
made in the previous section and obtain:  
\begin{equation}
\dfdm^{(b)}(k,q,a)=f_\dm(q,a) h(k,a_q) \jmath_\ell(k\omega_q(a,a_q))
\,,\label{eq:dfdmb} 
\end{equation}
where $a_q=q/\pcm$,
$\omega_q(x,y)=\int_{y}^{x} d\!a
(q/\epsilon(q,a))/\adot$ and
$\epsilon(q,a)=\sqrt{m^2a^2+q^2}$ is the comoving energy. Note that
$\omega_q(x,y)$ is the  comoving distance traversed by a particle with
comoving momentum $q$ between $x$ and $y$. $h(k,a)$ is the
perturbation to the trace of the the synchronous gauge spatial metric
in fourier space written as $\delta_{ij}\delta({\vec
  k})/3+\delta_{ij}h(k,a)+ ({\hat k}_i{\hat
  k}_j-\delta_{ij}/3)\alpha(k,a)$ \cite{ma95}.  

We have modified CMBfast \cite{seljak96} to incorporate the above
physics. The physics of how the decay affects the dark matter
perturbations is simple. The decay of DDM to DM provides a large
momentum to the DM particle. This implies that the mean free path of
DM is larger than in the standard case. Perturbations on scales
smaller than the ``mean free path'' of the DM particle cannot
survive.

How is this manifested in the Boltzmann equation? The solution of the
linearized collisionless Boltzmann equation (valid when the decay term
can be neglected) can be written as  
\begin{eqnarray}
\delta \rho(a_0) &=& (2\pi)^{-2} a_0^{-4} \int_0^{a_0} d\!a \int_0^{\infty}
d\!q \epsilon q^3 {df_0 \over dq} \left[ \right. \nonumber \\
&-&{h' \over 3} j_0(k\omega_q(a,0))+{2 \over
    3}\alpha' j_2(k\omega_q(a,0))\left.\right]\,,\label{eq:freestream1}
\end{eqnarray}
where $h'$ and $\alpha'$ denote derivatives with respect to scale
factor $a$. The free-streaming effect comes in through the effect of
the function 
$\omega_q$. We note that the $j_0$ and $j_2$ terms result from the
plane wave expansion of $\exp(\imath k\omega_q(a,0))$. In the limit of
$m \rightarrow 0$, we have a plane wave (which is how the photons
free-stream after decoupling). In the other limit of $m\rightarrow
\infty$, $j_0 \rightarrow 1$ and $j_2 \rightarrow 0$, and we obtain
the evolution of the CDM density perturbations. Similar expressions
were first derived by Bond and Efstathiou \cite{bond80}, and Brandenberger,
Kaiser and Turok \cite{brandenberger87}. On small scales, one may make
further approximations and 
write $\delta \rho(a_0) = (2\pi)^{-2} a_0^{-4}\int_0^{a_0} d\!a
(h'+\alpha')/3\int d\!q q^2 \epsilon f_0
d\ln(w_q)/d\ln(q)\cos(k\omega_q)$. Note that when
$k\omega_q(a,q/\pcm)$ gets large, we get no contributions to the
integral due to cancellations from rapid oscillations.
The power spectrum is thus suppressed.  

A point of interest in this discussion is that the quantity that
determines the damping of the power spectrum on small scales is not
the canonically defined mean free path 
$\int_{\tau}^{t_\mathrm{eq}} dt v(t)/a(t)$. Damping occurs when
$k\omega_q(a,q/\pcm) > 1$; very roughly, this condition picks out
scales smaller than $0.005 (Q/[\msun/\pc^3/(\kms)^3])^{-1/3}
\mpc$. 


What are the natural variables to describe this class of ``warm dark
matter'' models? We may take $\adec$ and $m/\pcm$ to be the two
variables. $\adec$ is the scale factor of the universe when the age is
equal to the lifetime $\tau$. The cosmological consequences depend
only on these two variables. 

If we specialize to the case of decays in the radiation dominated era,
then we will find that there is just one variable that adequately
describes the cosmological effects. For those decays that release EM
energy, decays must occur in the radiation dominated era in order to
not distort the CMB black body spectrum beyond the $10^{-4}$ level.

A physically motivated variable is phase space density of DM particles
in the early universe. We will adopt the definition of Hogan and
Dalcanton \cite{hogan00} who defined $Q=\rho/\langle v^2
\rangle$. For times much larger than the decay lifetime, we have the
exact relation  
\begin{equation}
Q=10^{-24} \left({m \over \pcm \adec}\right)^3 {\msun/\pc^3 \over
  (\kms)^3} \,. \label{eq:qprim}
\end{equation}
We will see that as a first approximation, the power spectrum only
depends on $Q$. For reference, we note that in the case of the slepton
decaying to the gravitino, $Q=2.1 \times 10^{-3} (2 \pcm/M)^3(M/\TeV)^{4.5}
\msun/\pc^3/(\kms)^3$. 

The net result on the power spectrum can be written in a manner
analogous to the massive neutrino case. The power-spectrum is
suppressed on scales smaller than $\lambda_c$. Here we provide
a simple fitting formula that encapsulates the basic features of the
power spectrum at the 25 \% level for lifetimes less than about a
year and $k < 25 h/\mpc$. 
\begin{eqnarray}
P_{\rm DM}(k) &=& P_{\rm CDM} \left({1 \over 2} 
\exp(-(k\lambda_c)^2/2)\right.\nonumber \\
&+&\left.{1\over 2}\left[1+(k\lambda_c)^3\right]^{-1}\right)^2\,,
\label{eq:modps}
\end{eqnarray}
where $\lambda_c = 0.0198 / Q^{0.275} \mpc$. At this
level of accuracy, we thus have a one parameter family of
models.

The above power spectrum should be compared to the power spectrum in
Warm Dark Matter (WDM)
models. Bode, Ostriker and Turok \cite{bode00} quote $P_{\rm WDM} =
P_{\rm CDM} (1+(\alpha k)^{2\nu})^{-10/\nu}$ based their analytic work
and the exact results of Ma \cite{ma96}, where $\nu=1.2$ and $\alpha=
0.048 h^{-1}\mpc\, 
(\Omega_{\rm   WDM}h^2/0.169)^{0.15}({\rm   keV}/m)^{1.15}$ for a
fermionic WDM with two internal degrees of freedom. Note that the
asymptotic power-law in the two models are different; the decay model
has comparatively more power due to its non-thermal
distribution. Looking at the variance $\sigma(M)$, we find that the
modified DM power spectrum 
with $\lambda_c=0.1 \mpc/h$ is a good fit to the variance function for
a 1 keV WDM model. Bode, Ostriker and Turok \cite{bode00} find that
the large scale 
structure constraints are met by a WDM model with 1 keV
particle. Thus, $\lambda_c \la 0.1 \mpc/h$ will also reproduce the
large scale structure we observe. For reference we observe that a keV
WDM implies $Q = 5 \times 10^{-4} \msun/\pc^3/(\kms)^3$. From the
approximate formula following Equation \ref{eq:modps}, we get a
similar $Q$ value for $\lambda_c = 0.1 \mpc/h$  and $h=0.65$. 

\section{Phase-space density constraints}
In warm dark matter models like those due to massive neutrinos, an
important constraint for late time cosmology results from the finite
phase space density \cite{tremaine79,hogan00}. For a warm dark matter 
fermionic particle which decouples when relativistic, the phase space
density frozen in is a Fermi-Dirac distribution. The maximum allowed
phase space density for a Fermi-Dirac distribution is given by
$h^{-3}/2$. The resulting gravitational distortions of the phase-space
sheets can never exceed this phase space limit for collisionless
particles. This is a direct consequence of
Louiville's theorem as applied to collision-less systems (also called
the Vlasov equation).   

For the decaying dark matter particle at hand, the maximum phase space
density, though finite, can be very large because of the $1/q$
term. We note that $q$ tends to a constant when the parent particle is
relativistic. 

The maximum phase space density argument is not the strongest
statement that can be made regarding the evolution of collision-less
systems. Consider a phase space distribution that is
Fermi-Dirac plus a delta-function at some small $q$ value. The maximum
phase space density argument would have nothing to say in this
case. However, if the total number of particles within the
delta-function spike is small, then we don't expect them to affect the
evolution of the collision-less system. 

Lynden-Bell \cite{lynden-bell67} showed that collision-less systems
have an infinity of conserved quantities. He labeled these
$M(>f)$, the mass of particles (or phase space cells) with densities
greater than a value $f$. There is however a problem in using $M(>f)$
to make statements about the evolution of a collision-less
system. Lets consider the case of a uniform density of particles
collapsing to form a galaxy. The observational information we have
about the galaxy is never about the fine-grained distribution of the
system. We only measure averaged quantities -- the coarse-grained
distribution. $M(>f)$ is not conserved for coarse-grained
distributions; it could increase or decrease. Tremaine \etal
\cite{tremaine86} proved a theorem using the Boltzmann H-functionals
that coarse-graining decreases these H-functionals (concave functions
of the distribution function, like the entropy). What about comparing
two coarse-grained systems? 

Dehnen \cite{dehnen05} (see also the early work by Mathur
\cite{mathur88})  recently proved that there exists a function -- the
excess mass function -- that always decreases as a result of
coarse-graining. The more coarse-grained, the smaller this excess mass
function. The excess mass function is very close in spirit to
Lynden-Bell's $M(>f)$. It is defined as: 
\begin{eqnarray}
D(f) \equiv 
\int d^3x d^3q (F({\vec x},{\vec q}) - f) \Theta(F({\vec x},{\vec q})-f)
\label{eq:excessmass}
\end{eqnarray}
Let us look at the two important features of this function in a little
more detail. First, for collisionless systems $D(f)$ is an
invariant. In a collisionless system, the phase space density is
conserved along a world-line (solution of the equations of
motion). Thus the total ``mass'' in elements with $F > f$ must be the
same (even though the global distribution of this mass could have
changed). Second, for collisionless systems, $D(f)$ decreases as a
result of coarse-graining. In order to see why this is so, let us
coarse-grain by dividing up our phase space into ``macro'' phase
space cells \cite{lynden-bell67}. Each 
macro-cell of phase space is made up of many (at least more than one)
micro-cells. Suppose that the macro-cell created in this way has $F
>f$. Now those micro-cells in this coarse-graining that had densities
larger than $f$ don't change $D(f)$. However, the micro-cells with
densities smaller than $f$ that get included in $D(f)$ (post
coarse-graining) do change $D(f)$. In fact, it can only decrease
$D(f)$ because the $F-f$ is negative for them. We can similarly argue
that a macro-cell created with $F<f$ also decreases $D(f)$. Thus
coarse-graining can only decrease $D(f)$.

This provides a natural way to constrain the inner cores of dark
matter halos. $D(f)$ for the halo must be smaller than the $D(f)$
calculated for the primordial distribution function for all
values of $f$. We note that this constraint is stronger than demanding
that the entropy of the halo be larger than the entropy of the
collection of particles that make up the galaxy halo in the early
universe. This may be ascertained once the change in entropy is
written as $-k\int_0^\infty df \Delta D(f)/f$ \cite{dehnen05} where
$\Delta D(f)$ is the change in the excess mass function.  

\section{Dark matter halos and their cores}

As an application of the above discussion consider a halo of DDM
particles that has a King profile. These profiles have a core of
almost constant density, a roughly $1/r^2$ density run and finally a
sharp drop in density to zero. The benefit of these profiles over the
usual isothermal profiles is that the King profiles have finite mass.  

The small galaxies in the local group have large $Q$ values. From a
compilation of Local group dSph galaxies by Mateo \cite{mateo98}, we
infer that the largest value of $Q$ is observed in Sculptor
with a core radius of 110 pc and 1-D stellar velocity dispersion of
6.6 km/s. Dalcanton and Hogan \cite{dalcanton00} show that one may
assign a $Q$ value of $2 \times 10^{-4} \msun/\pc^3/(\kms)^3$ to this
galaxy. They also assign $Q$ values to seven other
local group galaxies with measured stellar velocity dispersions
between 6 and 11 km/s and $M_v >  -14$. The $Q$ values range between
$10^{-5}$ and $10^{-4} \msun/\pc^3/(\kms)^3$.  

\begin{figure}[htbp]
\includegraphics[width=\columnwidth]{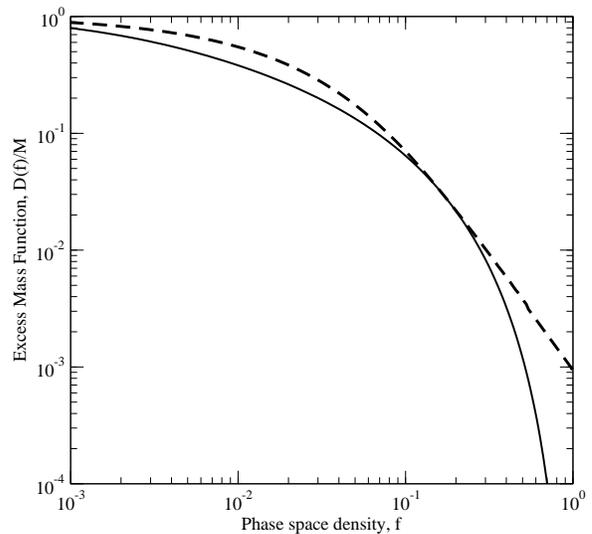}
    \caption{\label{fig:excessmass}
      The solid curve shows the excess mass function for a dark 
      matter halo with a mass of $2 \times 10^7 \msun$ and a King
      profile. The dashed curve shows the excess mass function for the
      primordial phase space distribution of dark matter from decays
      with $Q=2 \times 10^{-5} \msun/\pc^3 (\kms)^{-3}$. 
}
\end{figure}

We choose a King profile for the dark matter halo with a concentration
equal to 24 and central density set by the observed stellar profile core
and stellar velocity dispersion \cite{pryor90}. The total mass of the 
halo thus chosen is $2 \times 10^7 \msun$. 
We plot the excess mass function, normalized to the
total mass, for this profile in Figure \ref{fig:excessmass}. We also
plot the primordial excess mass function that is greater than, but
barely so, than the halo excess mass function.

The King density profile drops to zero exponentially. The drop is
sharp enough that for some small value of $f$, the King excess mass
function will always be larger than the primordial one. Therefore, in
the above comparison we only look at $f$ values such that $M(>f) <
0.99 M(>0)$. Any realistic profile will have a more benign behavior at
small $f$; indeed, the excess mass function constraint will impose
such behavior.  

The above exercise indicates that the decay parameters must be such
that $Q_{\rm prim}>2 \times 10^{-5} \msun/\pc^3 (\kms)^{-3}$. 
We note that if the halo of Sculptor is less massive than we have
assumed, then $Q_{\rm prim}$ will have to be larger. If the halo
is more massive, then $Q_{\rm prim}$ can be lower. In scenarios with
a small scale cut-off in the power spectrum, the second possibility is
more likely because the cut-off leads to a paucity of small mass halos 
\cite{bullock02}.

We also note that requiring the excess mass function of
the galaxy to be smaller than the fine-grained excess mass function in 
the early universe is a minimal constraint. The final effect on the
galaxy of a small primordial phase space density is likely more
complicated because of halo mergers.

An interesting exercise at this point is to compare the constraint
above to that obtained by demanding that the entropy of the collection
of particles that make up the galaxy can only increase. We define
entropy for N indistinguishable particles as   
$S=- kN \int d^3x d^3p f({\vec p},{\vec x}) [\ln(f({\vec p},{\vec
    x})h^3)-1]$. For dark matter from decays, this works out to
$kN\ln[2\pi e^{2-\gamma/2}m^4 Q^{-1} h^{-3}]$, which is very close to
the entropy for an ideal gas of particles with the same $Q$. A
comparison with the King profile entropy for Sculptor galaxy yields
the result that in order for the entropy to not decrease we must have
$Q_{\rm prim} > 10^{-5} \msun/\pc^3/(\kms)^3$, weaker than the
constraint obtained from the excess mass function analysis. 

The study by Dalcanton and Hogan \cite{dalcanton00} found that the
lowest mass halo core densities could be interpreted as resulting from
a finite primordial phase space density. However, they found no
compelling argument to attribute to the core density vs. velocity
dispersion they found over a wide dynamic range to the effect of a
small primordial phase space density. 

The effect of a finite $Q$ goes beyond just smoothing the inner cores
of small galaxies. Larger galaxies are a result of the merger of
smaller galaxies and hence the cumulative effects on the halo density
of larger dark matter halos could be substantial. Zentner and Bullock
\cite{zentner03} used semi-analytic arguments and the merger tree
formalism to show that a small phase space density can lead to a
lowered concentration for dark matter halos. For the fermionic warm
dark matter candidates, they calculate $c(M)$ for a given virial speed
and $Q$. We find that $c(M) \propto Q^{0.63}$ approximates their results
well for the range of $Q$ values of interest. The exact
relationship between $c(M)$ and $Q$ is more complicated in the present
model as compared to the warm dark model case. This is because the
constraint on $Q$ depends on the final phase space profile of the
halo. A more detailed study is required on this subject.

Why is this important? The concentration of $\Lambda$CDM halos as
obtained from fits to low surface brightness galaxies (LSBs) falls on
the low side. This might be due to two reasons. One, LSBs are extremely
strongly biased to forming in the lowest concentration halos. Two, CDM
is not the correct description of dark matter. As yet, we have no
compelling model advocating that the first reason is correct. Clearly,
the model  presented here has the right features to help explain the
observed low concentrations.

The small scale end of the CDM power spectrum has to contend with
another issue. CDM predicts a lot of sub-halos in the halos of the
kind that would host the Milky Way. However, we see almost a factor of
ten smaller number of galaxies \cite{moore99}. On the other hand CDM
substructure seems to be required to explain the strong lensing
anomalies \cite{kochanek04}. Plausible astrophysical solutions
\cite{bullock00} to this problem certainly exist, as do more exotic
ones.  Kamionkowski and Liddle   
\cite{kamionkowski99} advocated reducing small scale power to explain
the lack of  small galaxies in the halo of the Milky Way.  
Zentner and Bullock \cite{zentner03} looked at this  issue in detail
and worked out the substructure fraction for linear power spectra with
a small scale cut-off. Their semi-analytic results applied
to our model of dark matter from early decays, and the results of the
WDM simulations by Colin \etal \cite{colin00}, suggest that this kind
of dark matter will help alleviate the above mentioned discrepancy.    

Another problem that has frequently been discussed is the
observational case against cuspy halos that CDM simulations predict
\cite{moore94,flores94,kravtsov97,moore98}. Low surface brightness
and dwarf galaxies seem to be better fit with a density profile that
flattens towards the center (e.g.,
\cite{weldrake03,swaters03,simon04,spekkens05}). Such evidence for a
core in higher mass galaxies is lacking. However, it is also true that
that the larger galaxies are  dominated by baryons in the centers
making it harder to detect a core if it were present. Dark matter from
decay of WIMPs can introduce a core in the smallest mass halos $\sim
10^7 \msun$ if $Q_{\rm prim}$ is small enough, $Q_{\rm prim} \sim
10^{-4} \msun/\pc^3/(\kms)^3$, and reduce the
concentration parameter of larger halos. A model with such low $Q$
values would produce a cut-off in the power spectrum given by
$\lambda_c=0.25 \mpc$. Is this consistent with having the universe
reionize fully at a redshift of six? We explore this in the next
section.  

\section{Reionization}
Reionization provides a stringent constraint on the small scale power
spectrum. The universe is reionized to a redshift of 6 and so there
should at least be enough small scale power to do that. The recent
WMAP results hint \cite{kogut03} that the optical depth to Thomson  
scattering might be large ($\sim 0.1$) indicating early
reionzation. If true, this would have dramatic implications for the
parameter space of dark matter from decays.

Barkana, Haiman and Ostriker \cite{barkana01} considered the
constraints on warm dark matter candidate from cosmological
reionization. Their detailed considerations led to the conclusion that
WDM with masses larger than about a keV could reionize the universe at
redshift six. 

Here we perform a simple analysis to understand the effect of the
decay on reionization using the semi-analytic models of Haiman and
Holder \cite{haiman03}. The ionized fraction is written as 
\begin{eqnarray}
&&F_{\rm HII}(z)=\rho_{\rm b}(z) \int_{\infty}^{z}dz^{\prime} 
\epsilon \left[ \frac{dF_{\rm coll,Ib}}{dz} (z^{\prime}) \right. 
\nonumber \\ 
&+& \left. \left(1-F_{\rm HII}(z^\prime)\right)
\frac{dF_{\rm coll,Ib}}{dz} (z^{\prime}) \right] {\tilde V}_{\rm
  HII}(z',z)\,, 
\label{eq:filling}
\end{eqnarray} 
where $\rho_{\rm b}$ is the average baryon density and ${\tilde
  V}_{\rm HII}(z',z)\epsilon M$ is the volume of region ionized at
  redshift $z'$ in a halo of mass $M$. 
The  subscripts Ia and Ib \cite{haiman03} refer to different
virial temperature ranges. Type Ia halos have temperatures between
$10^4$ K and $2\times 10^5$ K, while Ib halos have temperatures higher
than $2\times 10^5$ K. In principle there is also a contribution from
halos with virial temperatures smaller than $10^4$ K. This aspect has
gotten a lot of attention recently because it could be very important
for an early reionization epoch. We are neglecting this contribution
here to make a better comparison between CDM and DM from decays. In
models with suppressed small scale power, this contribution
is small. 

The reionization model as written above has two free parameters:
$C_{\rm HII}$, $\epsilon$. $C_{\rm HII}$ is the clumping factor for
the IGM predicted by CDM simulations to be of order 10 at high
redshifts. In the decaying DM
scenario, the clumping factor may be smaller, thus impeding
recombinations. For our simple analysis, we will take $C_{\rm HII}=15$
for CDM and $C_{\rm   HII}=10$ for the DM from decays. A robust
calculation will require input from n-body simulations. 

The efficiency parameter is the product of the average (over IMF)
number of ionizing photons over the lifetime of the source, the escape
fraction of ionizing photons and the fraction of baryons converted to
sources (stars). We take the efficiency to be 100. Given standard IMFs
for metal--rich stars, we expect 4000 ionizing photons per
baryon. Thus an efficiency of 100 implies escape fraction times
fraction of baryons in stars of 2.5\% in keeping with what is observed
in the local universe. 

With the above inputs, the $\Lambda$CDM
model gives a optical depth of 0.091. For $\lambda_c=0.1 \mpc/h$, we get
an optical depth of 0.061, while for  $\lambda_c=0.3 \mpc/h$ we get
0.035. However, for $\lambda_c=1 \mpc/h$ we get an optical depth of
0.017, inconsistent with an ionized $z<6$ universe. 

\begin{figure}[htbp]
  \includegraphics[width=\columnwidth]{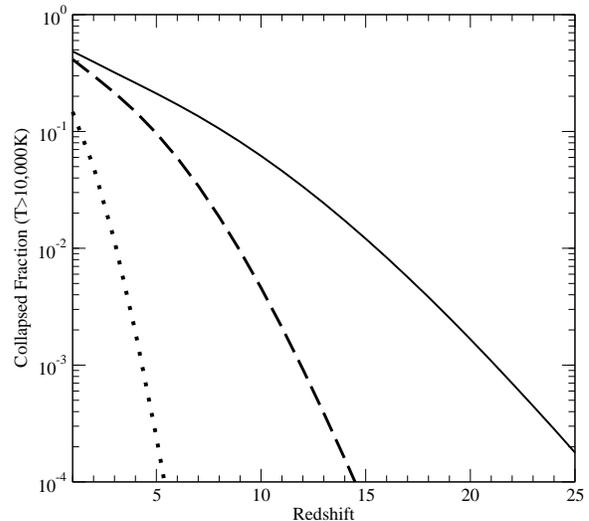}
    \caption{\label{fig:fcoll}
      The curves show the collapsed fractions of halos with
      virial temperatures larger than $10^4$ K. Below this
      temperature, the gas can only cool by molecular hydrogen. The
      solid curve corresponds to a standard $\Lambda$CDM model with no
      tilt. The dashed curve corresponds to the same model but with a
      linear power spectrum modified according to Equation
      \ref{eq:modps} with $\lambda_c=0.1 \mpc/h$. The dotted curve has 
      $\lambda_c=1 \mpc/h$. 
}
\end{figure}

The calculations of this section show that in order to be able to
reionize the universe by redshift six, we expect that $\lambda_c \la
0.3 \mpc/h$, which implies  $Q_{\rm prim} \ga 10^{-5}
\msun/\pc^3/(\kms)^3$. 
The constraint is similar to the one we found in the last section by 
considering the phase space constraint from local group galaxies. If
the universe is reionized at higher redshifts than six, this constraint
will tighten considerably. 

How reliable are the above calculations? Along with the simplistic
modeling of the reionization process, one of the weak
points of the above calculation is the calculation of the collapsed
fraction. We calculate $\sigma^2(M)$ using the modified linear power
spectrum and use the extended Press-Schechter \cite{press74} scheme of
Sheth and Tormen \cite{sheth01} to calculate the mass function of dark
matter halos. The results of this calculation for the 
collapsed fraction of halos with $T > 10^4$ K are shown in Figure
\ref{fig:fcoll}. It is unclear that such a calculation will yield
accurate results on small scales where the suppression in power
becomes important and where structure formation is not fully
hierarchical. 

\section{Mixed models}
In scenarios of the kind we have discussed above, we may expect some
of the DM to be produced during reheating after inflation. Let us take
the case of the gravitino DM. If the reheating temperature is around
$10^{10}$ GeV then we expect comparable contribution to the total
number density of gravitinos today from the decay and the reheating
phase (see Roszkowski and Austri \cite{roszkowski04} for recent work
on early universe constraints on these models). The gravitinos from
the reheating phase would behave as Cold Dark Matter particles. 

\begin{figure}[htbp]
\includegraphics[width=\columnwidth]{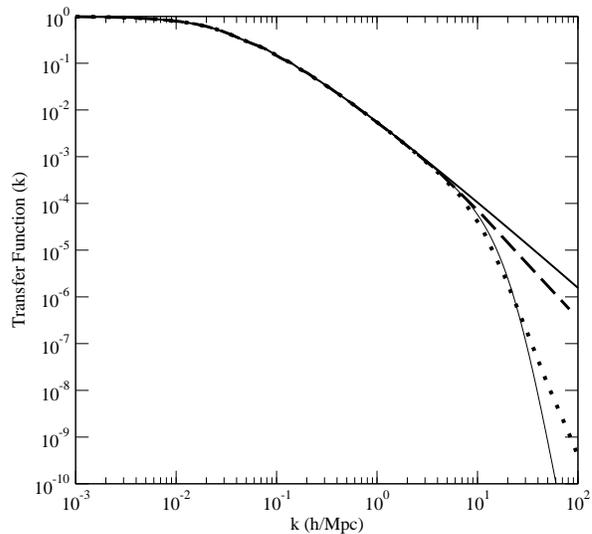}
    \caption{\label{fig:fdecay}
      The curves show the power spectrum for different
      values of $f$, the fraction of dark matter today that came from
      decays. The solid curve shows the $f=0$ case (CDM). The dashed
      curve shows the $f=0.5$ case while the dotted curve shows the
      $f=1$ case. It is clear that the suppression on small scales is
      much reduced for the $f=0.5$ case. For comparison, we also plot
      (see thin solid curve) the transfer function for a 1 keV Warm
       Dark Matter model.
}
\end{figure}

The main impact of these ``mixed models'' is to soften the small scale 
effects. This was considered in detail by Profumo \etal
\cite{profumo04} for the charged particle decay scenario. We write
the dark matter density today as $\rho_{\rm DM}=f\rho_{\rm
  DEC}+(1-f)\rho_{\rm CDM}$ where $f$ is the fraction of the dark
matter that results from the decay. The $z=0$
transfer function for the total dark matter is shown in Figure
\ref{fig:fdecay}. Note that the suppression on small scales is 
reduced as expected. The form of the transfer function cannot be
trivially obtained from the transfer function for the $f=0$ and $f=1$ 
cases. 


\section{Discussion}
The phenomenology of DDM models is rich and much work remains to be
done. To make robust contact with data on small scales we need large
scale structure simulations and semi-analytic models. A few detailed
simulations in this regard have been run for models with keV WDM
particle \cite{bode00,colin00}, with linear power spectrum cut-off on
small scales \cite{white00}, and to understand the dependence of the
power spectrum on substructure \cite{eke01}. 

We looked at some beneficial features of these models in the previous
sections and also the constraints on them. What are the other 
constraints? Narayanan \etal \cite{narayanan00} show that in the WDM 
scenario, masses smaller than 750 eV are disfavored by the Ly-$\alpha$
data. Recent work by Viel \etal \cite{viel05}
peg the lower limit on the WDM mass at 550 eV at $2\sigma$. 
If we naively compare the variance $\sigma(M)$ in the WDM and
the present model, then this may be turned into a constraint on the
phase space density parameter (see Equation \ref{eq:qprim}), $Q >
10^{-4} \msun/\pc^3/(\kms)^3$ corresponding to WDM masses larger than
750 eV, and $Q > 5 \times 10^{-5} \msun/\pc^3/(\kms)^3$ corresponding
to WDM masses larger than 550 eV.

Thus, we may summarize the present constraints on models where the
bulk of dark matter today results from decays with a lifetime of about
a year or smaller as $Q \ga 10^{-4} \msun/\pc^3/(\kms)^3$. 
Future comparisons with data will require large simulations to
understand  structure formation in these models; it does not proceed
in a bottom-up hierarchical manner \cite{white00,bode00,colin00}.

Can we push down to 100 kpc or even 10 kpc cut-off scales? Strong
lensing flux anomaly technique is a sensitive probe of the
amount of substructure \cite{dalal02}. Robust limits
from this technique will require detailed theoretical understanding of
issues such as the anisotropic distribution of substructure in dark
matter halos \cite{zentner05}. Weak lensing constraints
from future surveys is another promising avenue to learn more about
dark matter on these small scales \cite{dolney04}. 

\section{Conclusions}
In this paper, we have explored the important cosmological
consequences of dark matter from early decays -- the cut-off in
the power spectrum on small scales, and the limit on the phase
space density of dark matter in halos. The phenomenology is rich and
there are multiple ways to search for the effect of such dark matter
on structure formation. We pointed out that these models can suppress
small scale substructure, create constant density cores in small mass
halos due to the phase space constraint and reduce the concentration
of larger mass halos. The models that give rise to these early
decays in supersymmetric theories are natural and they inherit the
correct cosmological abundance from the WIMPs that decay into
them. Future observations of structure on small scales may be able to
distinguish between cold dark matter and the dark matter from early
decays.  

\section{Acknowledgments}
I thank Jose Cembranos, Jonathan Feng, Fumihiro Takayama and Andrew 
Zentner for discussions. I thank James Bullock for many discussions
about the problems on small scales, Neal Dalal for discussions during 
the early phase of this project and Arvind Rajaraman for convincing me
about the usefulness of the excess mass function. We acknowledge the
use of CMBfast code \cite{seljak96} for this work.

\bibliography{cmb3}

\end{document}